# Large-scale Group Brainstorming using Conversational Swarm Intelligence (CSI) versus Traditional Chat

Louis Rosenberg[1][a], Hans Schumann[1], Christopher Dishop[2], Gregg Willcox[1], Anita Woolley[2][b], and Ganesh Mani[2][c]

[1]*Unanimous AI, 2200 North George Mason Dr, Arlington, VA, USA*
[2]*Carnegie Mellon University, Pittsburgh, Pennsylvania, USA*
*Louis@Unanimous.ai, cdishop@andrew.cmu.edu, ganeshm@andrew.cmu.edu, awoolley@andrew.cmu.edu*

Abstract: Conversational Swarm Intelligence (CSI) is an AI-facilitated method for enabling real-time conversational deliberations and prioritizations among networked human groups of potentially unlimited size. Based on the biological principle of Swarm Intelligence and modelled on the decision-making dynamics of fish schools, CSI has been shown in prior studies to amplify group intelligence, increase group participation, and facilitate productive collaboration among hundreds of participants at once. It works by dividing a large population into a set of small subgroups that are woven together by real-time AI agents called Conversational Surrogates. The present study focuses on the use of a CSI platform called Thinkscape to enable real-time brainstorming and prioritization among groups of 75 networked users. The study employed a variant of a common brainstorming intervention called an Alternative Use Task (AUT) and was designed to compare through subjective feedback, the experience of participants brainstorming using a CSI structure vs brainstorming in a single large chat room. This comparison revealed that participants significantly preferred brainstorming with the CSI structure and reported that it felt (i) more collaborative, (ii) more productive, and (iii) was better at surfacing quality answers. In addition, participants using the CSI structure reported (iv) feeling more ownership and more buy-in in the final answers the group converged on and (v) reported feeling more heard as compared to brainstorming in a traditional text chat environment. Overall, the results suggest that CSI is a very promising AI-facilitated method for brainstorming and prioritization among large-scale, networked human groups.

## 1 INTRODUCTION

Humans are not the only species that deliberate in groups to reach decisions. Fish schools, bird flocks, and bee swarms are well known examples of natural groups that reach rapid decisions on life-or-death issues. Biologists refer to this collaborative decision-making process as Swarm Intelligence (SI) and it enables many social organisms to make decisions that are significantly smarter than the individuals could achieve on their own (Krause, et. al, 2010).

Artificial Swarm Intelligence (ASI) is a novel technology developed in 2014 to enable networked groups to quickly reach collaborative decisions as real-time systems modelled after biological swarms (Rosenberg, 2015). ASI has been shown to amplify the accuracy of group decisions across a wide range of applications, from financial projections and sales forecasting to business prioritization and medical diagnosis (Askay, et. al., 2019. Rosenberg, 2016).

While ASI is an effective technology, it requires that participants choose among a pre-defined set of options. This works for certain applications such as collaboratively prioritizing predefined sets of options or making numerical estimations or forecasts, but it is not useful for open-ended discussions, brainstorms, deliberations, or debates. To address this, a next-generation technology called Conversational Swarm

Intelligence (CSI) was developed in 2023 that combines the principles of Swarm AI with the power of large language models (Rosenberg, et al., 2023).

The goal of CSI is to enable large, networked human groups (50 to 500 people) to hold thoughtful conversational deliberations in real-time that rapidly converge on optimal solutions based on the combined knowledge, views, and opinions of the participants. To make this viable, researchers had to overcome several fundamental barriers related to basic human conversations. First and foremost, research shows that deliberative conversations are most effective in small groups of only 4 to 7 individuals and rapidly lose effectiveness with increasing size (Cooney, et. al., 2020). With additional members, all participants are afforded less and less airtime to express their views, and longer and longer wait times to respond to others. When a group reaches sizes larger than 10 to 12 people, it ceases to be a true deliberation and devolves into a series of monologues.

So how can a technology enable hundreds of people hold a productive real-time deliberation in which participants brainstorm solutions, build on the ideas of others, debate options and alternatives, and converge on solutions? To overcome this barrier, CSI takes its core inspiration from the decision-making dynamics of large fish schools. That is because large schools have thousands of members and provide an interesting analog to human organizations. Consider the image below which shows a large school facing three simultaneous threats that require a rapid and effective response:

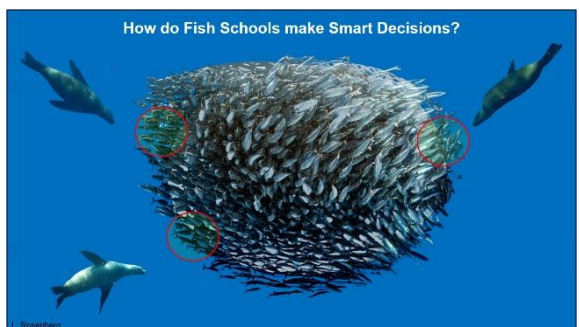

Figure 1. Fish School facing simultaneous threats.

In the figure above, three predators approach the school, creating a complex problem of life-or-death significance. Like many human organizations, the members of the school all have limited information. In fact, only three small pockets of fish are aware of any predators (the circled areas above). In fact, the vast majority of fish are unaware of any predator and those in the circled areas are only aware of one. So how can this large real-time system quickly make an optimal decision of which direction the school should move?

Fish schools use a unique form of communication among neighbouring individuals. Each fish has a specialized organ on the sides of their bodies called a lateral line that detects faint pressure and vibration changes in the water as the adjacent fish adjust their direction and speed. This enables small subgroups of neighbors to "deliberate" in real-time, establishing a multi-direction tug-o-war that converge on a direction that small subgroup of fish will go. And because each subgroup of neighboring fish overlaps other small subgroups, information quickly propagates across the full population. This enables an emergent property that biologists call Swarm Intelligence, and it allows thousands of individuals, each with a limited view of the world around them, to rapidly converge on unified decisions that are critical for survival (Parish, et. al., 2002. Rosenberg, et. al., 2023). Figure 2 below shows information propagating leading to an efficient and effective collective decision.

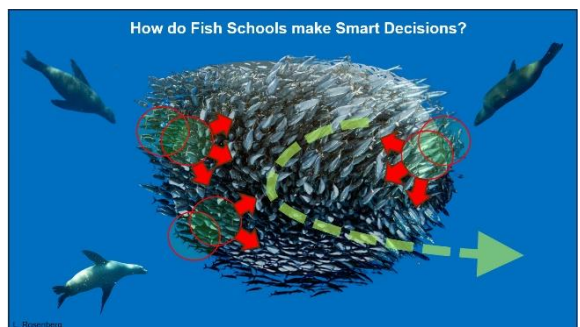

Figure 2. Swarm Intelligence enables optimized decisions.

CSI technology takes this natural process and emulates the dynamics by breaking large human groups into a network of overlapping subgroups, each with 4 to 7 members, as that size enables optimal conversational deliberation. Unfortunately, there is one more barrier that must be overcome – unlike fish, humans cannot participate effectively in overlapping subgroups (i.e. we did not evolve to participate multiple real-time conversations at once).

This is commonly called the Cocktail Party Problem – if you engage in a conversation with a small group at a party and get interested in what a neighboring group is discussing, you immediately lose focus on the original group (Bronkhorst, 2000). So how can hundreds of individuals hold a single conversation through overlapping subgroups?

To overcome this problem, CSI uses novel artificial agents called *"Conversational Surrogates"* that are powered by Large Language Models (LLMs) and enable the real-time overlap among deliberating groups (Rosenberg, 2023). Specifically, CSI breaks a large population into a series of parallel subgroups such that an LLM-powered surrogate agent is placed in each subgroup and tasked with observing the deliberation in that group, distilling the salient content, and passing critical ideas, insights, opinions and perspectives to other subgroups where that subgroup's local surrogate agent will express those points as a natural dialog within their ongoing conversation. With agents in all subgroups continuously observing insights and passing them to surrogate agents in other rooms, the full population is woven together into a single conversation in which ideas emerge and spread with high efficiency, along with arguments for and against those ideas. Using this novel architecture, 50, 500 or even 5,000 people can hold a real-time conversation in which they brainstorm ideas, debate alternatives, prioritize options and converge on solutions.

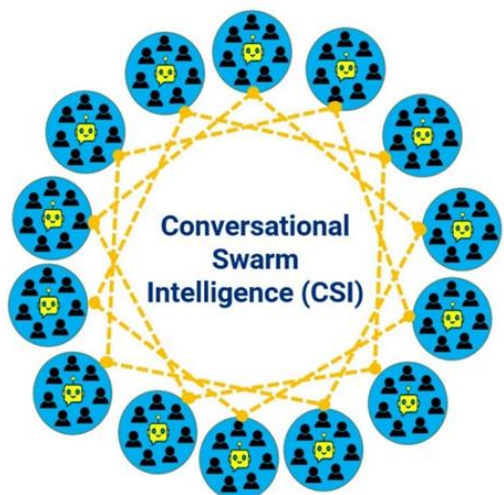

Figure 3. Conversational Swarm Intelligence Architecture

An example CSI architecture is shown in Figure 3 above in which a group of 98 people are divided into a network of 14 subgroups, each with 7 human users and one artificial agent. While the image implies that each subgroup can only pass information to two other subgroups in the network, the model employed enabled insights to pass from any subgroup to any other (i.e., a fully connected network). A unique matchmaking subsystem is used that that tracks (i) which groups have a new idea or insight that is ready to pass on to others, (ii) which groups have not received insights in a threshold amount of time and are ready to receive another, and (iii) which of the available ideas or insights would maximally change the receiving group based on what that group has discussed thus far.

In this way, CSI emulates the propagation of information present in fish schools but does so in a significantly more efficient manner. While schools can only pass insights among neighboring subgroups, CSI passes insights to any place in the network, selecting subgroups that are (a) ready to receive a new insight and (b) will be maximally challenged by the insight, opinion, or rationale received based on a real-time assessment of its local conversation thus far.

By facilitating large, networked populations to debate complex issues in real-time, CSI enables individuals with a wide range of knowledge, wisdom, and insights to collaboratively deliberate on broad, open-ended problems. And because every assertion expressed by every participant is identified and stored in a real-time taxonomy database by the CSI system, the system can immediately produce detailed forensic reports that reveal how each decision was reached, including a complete assessment of every idea raised, the reasons that support and reject each ideas, and impact each idea or reason had on others to sway the group towards a maximally supported solution.

In addition, CSI solves common biasing problems that drive deliberating groups to non-optimal answers. For example, groups can be overly impacted by individuals with strong personalities, with high rank within an organization, or who express ideas very early in a deliberation. This is mitigated by the CSI structure because points raised by a strong personality, a high-ranking individual, or an early talker in the deliberation only impact a small local subgroup. For those points to gain traction across the full population, they must stand on their own merits: either discussed organically in multiple subgroups or passed into subgroups by surrogate agents. Ideas that are passed into a group and significantly impact that group are more likely to pass to other groups, thus enabling strong insights to propagate quickly.

The effectiveness of CSI has been researched in a handful of recent studies. In one study conducted at Carnegie Mellon in 2023, groups of 48 participants were tasked with debating the future impact of AI on jobs using a CSI platform called Thinkscape™. The participants using CSI contributed 51% more content ($p<0.001$) compared to those using standard centralized chat. In addition, CSI showed 37% less difference in contribution between the most vocal and least vocal users, indicating that CSI fosters more balanced deliberations. (Rosenberg, et. al., 2023).

In another recent study, groups of 35 individuals were tasked with taking a standardized IQ test, either as individuals on a survey, as a "crowd" by taking the aggregation of surveys, or as a conversational swarm inside the CSI-powered Thinkscape platform. The groups of randomly selected participants using CSI averaged a collective of score 128 on the IQ test when working together as conversational swarm intelligence, significantly outperforming both the average individual participant (IQ 100, p<0.001) and a groupwise statistical aggregation across groupings of 35 individual tests (IQ 115, p<0.01). In addition, the score of 128 IQ achieved by the average CSI group placed its performance in the 97th percentile of individual IQ test takers, achieving "gifted" status by most metrics (Rosenberg, et. al. 2024).

While prior studies have shown that large groups using CSI (i) increase conversational participation, (ii) foster more balanced dialog among participants, and amplify collective intelligence compared to traditional methods, no prior study has explored the ability of large groups to brainstorm collaboratively and converge on a set of prioritized solutions in real-time. The following study aimed to test brainstorming among groups of approximately 75 individuals and assess their comparative perceptions of brainstorming with CSI versus brainstorming within a single large group in a traditional online chat platform.

## 2 BRAINSTORMING STUDY

To assess if large networked human groups can hold real-time brainstorming conversations using a CSI structure and converge on a small set of maximally supported solutions, two sets of approximately 75 people (sourced from a commercial sample provider) were assembled in the text-based Thinkscape platform and tasked with a collaborative brainstorming problem. As a baseline, the same groups we also assembled in a single large text-based chatroom of similar real-time functionality to Discord, Slack, Google Chat, Microsoft Teams and other commercial room-based chat environments.

The brainstorming task used was a modified version of a typical Alternative Use Task (AUT) that is given to assess creative abilities in individuals and/or groups (Habib, et. al, 2024; Guilford, 1967). In this case, two alternative use tasks were devised – a first task which asked groups to imagine they work for a large company that has been stuck with a significant inventory of traffic cones. Their task is to come up with as many alternative uses of traffic cones as possible (unrelated to traffic) that could be viable products sold the fictional company and to identify the best ideas among the proposed alternatives. The second task was structured the same way, but the item that the fictional company had in inventory were toilet plungers.

The protocol for the first group of 75 individuals was to first brainstorm the traffic cone AUT task first in a single large chat room and then brainstorm the toilet plunger AUT task in a CSI structure in which the 75 individuals were broken up into approximately 15 subgroups of 5 individuals, each sub-group including one AI agent (i.e., conversational surrogate) that participated in the local conversation by sharing ideas received from other subgroups. The second group of 75 performed the same protocol, but brainstormed traffic cones first in the CSI structure, then brainstormed toilet plungers second in a standard large chat room structure. At the conclusion of the intervention, both groups were given a survey in which they were asked a set of subjective judgment questions to compare each brainstorming experience, the single large room versus the CSI structure.

For clarity, when using CSI, each participant was only able to converse with the other 4 members of their subgroup and with the AI agent. The AI agents did not introduce any AI generated ideas or opinions into the local conversations – they only passed and received conversational ideas and opinions from other subgroups, weaving the set of 15 subgroups into a single conversion which individuals could build upon the ideas of others in alternate groups and/or share justifications in support or opposition to ideas across subgroups. In all trials, participants were given 12 minutes to complete each AUT brainstorm task.

## 3 DATA AND ANALYSIS

Each of the two groups of 75 participants took part in a 30-minute session in which they performed two AUT brainstorms for 12 minutes each (one using CSI and one in a traditional chat room) and then individually completed a subjective feedback survey to compare the two experiences. The questions asked on the survey were as follows:

- Which method felt more productive?
- Which method made you feel more heard?
- Which method felt more collaborative?
- Which method was surfaced better answers?
- Which method made you feel more buy-in?
- Which method made you feel more ownership?
- Which method did you prefer overall?

The only substantive difference between the two groups of participants was that Group 1 brainstormed in a standard chat room first, then used CSI, while the participants of Groups 2 brainstormed using CSI first and then used the standard chat room. This was to mitigate ordering effects on the subjective feedback.

In total we collected 147 surveys, each comparing brainstorming and prioritization using a CSI structure versus a traditional chat room. In the CSI structure, the 75 individuals brainstormed by being divided into 15 subgroups of 5 people, each subgroup including an AI agent that participated in their local conversation to link all the subgroups together. In the standard chat room structure, all 75 people were able to see the ideas of everyone else and respond to the full group.

The results were highly conclusive, showing that a significant majority of the 147 survey-responding participants preferred the CSI structure to the standard chat room structure in all seven questions asked. To assess if these results were statistically significant, a one-proportion z-test was performed on each question in the surveys to test if the results showed statistically significant evidence that more people preferred one method over the other. Because multiple statistical tests were run, we used a Bonferroni adjustment to determine significance at the 1% alpha level and needed to observe a p-value<0.01/7=0.0014 for each of the 7 questions tested. This level of significance was observed in each of the seven questions, meaning we can conclude with 99% confidence that participants preferred the CSI platform (Thinkscape) for brainstorming and prioritization as compared to traditional text chat.

## 4 RESULTS

The segmented bar chart in Figure 4 below shows the proportion of survey respondents that preferred either Thinkscape or Standard Chat when answering each of the feedback questions.

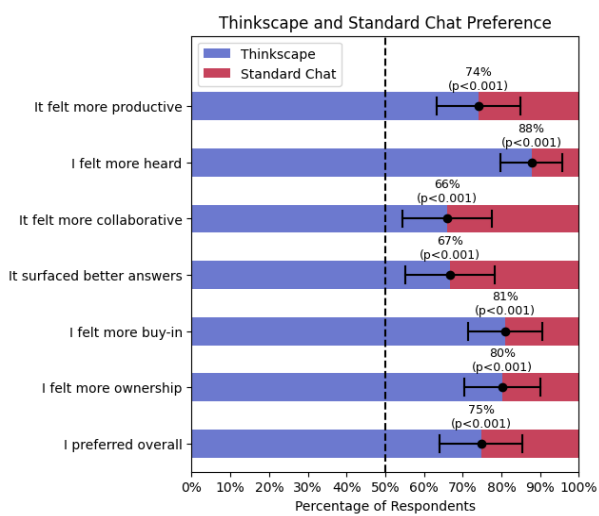

Figure 4. Subjective Feedback Results with Error Bars.

We can see in Figure 4 that a significant majority of participants preferred Thinkscape with respect to all seven of the feedback questions, the support ranging between 66% and 88%, with 75% of respondents preferring Thinkscape overall. Each question in Figure 4 also shows error-bars reflecting a 99% Bonferroni-adjusted confidence interval estimating the true proportion of all participants who would prefer Thinkscape over a Standard Chat. None of the confidence intervals overlap the 50% dotted line, demonstrating statistical significance in our findings that Thinkscape is the preferred method.

## 5 CONCLUSIONS

The results of this study are promising, showing that groups of 75 individuals can successfully brainstorm and prioritize in real-time using a text-based CSI platform. The results further show that participants significantly preferred the CSI structure (which used AI agents to weave together conversations among a large number of small subgroups) over the traditional structure of a single chatroom. In particular, they found the CSI structure to be more productive, more collaborative, and better at surfacing quality answers. In addition, over 80% of participants reported feeling *"more heard"* during the deliberation and came away feeling *"more ownership"* and *"more buy-in"* with respect to the resulting answers than in a traditional real-time chat environment.

Future studies into CSI should aim to evaluate collaborative brainstorming and prioritization among significantly larger groups, aiming to validate usage among hundreds or even thousands of individuals. Considering that the average Fortune 1000 company has over 30,000 employees, the ability to engage very large groups in real-time discussions, brainstorms, debates, evaluations, and prioritizations could be a powerful method for capturing feedback, fostering cross-pollination of ideas and insights, optimizing forecasts, amplifying intelligence, and promoting buy-in and ownership across large organizations.

Future studies should also validate the value of CSI in voice chat and video conferencing environments. And finally, future studies should explore the unique usefulness of CSI in vertical applications in which enabling thoughtful deliberation at scale is desired but traditionally hard to achieve. Examples of high-value applications include enterprise collaboration, citizen assemblies, deliberative civic engagement, employee feedback, big science, and consumer research.